\documentclass[aps,prl,showpacs]{revtex4}
\usepackage[latin1]{inputenc}
\usepackage{graphicx,amsmath,amsfonts,amssymb}
\usepackage{color}
\newcommand{\be}{\begin{equation}}
\newcommand{\ee}{\end{equation}}

\usepackage{ae}
\usepackage{float}

\begin{document}

\title{Position-momentum correlations in matter waves double-slit experiment}
\author{J. S. M. Neto, I. G. da Paz\footnote{Corresponding author.}}
\affiliation{ Departamento de F\'{\i}sica, Universidade Federal do
Piau\'{\i}, Campus Ministro Petr\^{o}nio Portela, CEP 64049-550,
Teresina, PI, Brazil.}

\author{L. A. Cabral}
\affiliation{Curso de F\'{\i}sica, Universidade Federal do
Tocantins, Caixa Postal 132, CEP 77804-970, Aragua\'{\i}na, TO, Brazil.}

\begin{abstract}
We present a treatment of the double-slit interference of
matter-waves represented by Gaussian wavepackets. The interference
pattern is modelled with Green's function propagator which
emphasizes the coordinate correlations and phases. We explore the
connection between phases and position-momentum correlations in the
intensity, visibility and predictability of the wavepackets
interference. This formulation will indicate some aspects that can
be useful for theoretical and experimental treatment of particles,
atoms or molecules interferometry.

\end{abstract}


\maketitle

\section{Introduction}
The double-slit experiment illustrates the essential mystery of quantum
mechanics \cite{Faynman}. Under different circumstances, the same
physical system can exhibit either a particle-like or a wave-like
behaviour, otherwise known as wave-particle duality \cite{Bohr}.
Double-slit experiments with matter waves were performed by
M\"{o}llenstedt and J\"{o}sson for electrons \cite{Jonsson}, by
Zeilinger et al. for neutrons \cite{Zeilinger1}, by Carnal
and Mlynek for atoms \cite{Carnal}, by Sch\"{o}llkopf and Toennies
for small molecules \cite{Toennies} and by Zeilinger et al.
for macromolecules \cite{Zeilinger2}.

Position-momentum correlations have been studied and interpreted in
some textbooks, while the most treated example is the simple
Gaussian or minimum-uncertainty wavepacket solution for the
Schrödinger equation for a free particle. Such wavepacket presents
no position-momentum correlations at $t=0$ which appear only with
the passage of time \cite{Bohm,Saxon}. How the phases of the wave
function influence the existence of position-momentum correlations
is also explained in Ref. \cite{Bohm}. Posteriorly, it was shown
that squeezed states or linear combination of Gaussian states can
exhibit initial correlations, i.e., correlations that not depend on
the time evolution \cite{Robinett, Riahi,Dodonov, Campos}.

The qualitative changes in the interference pattern as a function of
the increasing in the position-momentum correlations was studied in
Ref. \cite{Carol}. In addition, it was shown that the Gouy phase of
matter waves is directly related to the position-momentum
correlations, as studied by the first time in Refs.
\cite{Paz1,Paz2}. The Gouy phase of matter waves was experimentally
observed in different systems, such as Bose-Einstein condensates
\cite{cond}, electron vortex beams \cite{elec2}, and astigmatic
electron matter waves using in-line holography \cite{elec1}.  More
recently, it was observed that the position-momentum correlations
can provide further insight into the formation of above-threshold
ionization (ATI) spectra in the electron-ion scattering in strong
laser fields \cite{Kull}.

In this work, we use the previously developed ideas on
position-momentum correlations to analyze the Gaussian features of
the wavepacket and the interference pattern, as well as the
wave-like and particle-like behavior, in double-slit experiment with
matter waves. Before reaching the double-slit setup, the particle is
represented by a simple Gaussian wavepacket and, after the
double-slit apparatus, the particle is represented by a linear
combination of two identical Gaussian wavepackets coming from the
two slits. After the double-slit, the position and momentum of the
particle will be correlated even if the time evolution from the
source to the double-slit is zero. The correlations will be changed
by the evolution, enabling us to extract some information about the
interference pattern.

In section II we present the model for the double-slit experiment
considering that the matter wave propagates the time $t$ from the
source to the double-slit and the time $\tau$ from the double-slit
to the screen. Further, we calculate the wave functions for the
passage through each slit using the Green's function for the free
particle. In section III, we calculate the position-momentum
correlations and the generalized Robertson-Schrödinger uncertainty
relation for the state that is a linear combination of the states
which passed through each slit. In section IV, we calculate the
intensity, visibility and predictability to analyze the interference
pattern in terms of the knowledge of the position-momentum correlations .

\section{Double-Slit Experiment}\label{model}

In this section we return to the double-slit experiment and analyze
the effect of the position-momentum correlations in the interference
pattern. We consider that a coherent Gaussian wavepacket of initial
width $\sigma_{0}$ propagates during a time $t$ before arriving at
a double-slit that divides it into two Gaussian wavepackets. After
the double-slit, the two wavepackets propagate during a time $\tau$
until they reach the detection screen, where they are recombined
and the interference pattern is observed as a function of the
transverse coordinate $x$. As we will see, the number of
interference fringes and its quality are dramatically influenced by
the propagation times $t$ and $\tau$. In particular, there is a value
of time $t_{max}(\tau)$ for which the number of fringes tends to be
minimum. This value of time corresponds to a maximum separation of
the wavepackets on the screen and it is associated with one
maximum of the position-momentum correlations. On the other hand, if
the source of particles is positioned in such a way that, before arriving the screen, the
particles travel during a time interval which is not close to
$t_{max}(\tau)$, the number of interference fringes and its quality are increased significantly.

The wavefunction at the time when the wave passes through the slit
$1(+)$ or the slit $2(-)$ is given by

\begin{equation}
\psi_{1,2}(x,t,\tau) =\int_{-\infty}^{+\infty}
dx_{j}\int_{-\infty}^{+\infty}dx_{i}G_{2}(x,t+\tau;x_{j},t)F(x_{j}\pm
d/2)G_{1}(x_{j},t;x_{i},0)\psi_{0}(x_{i}),
\end{equation}
where

\begin{equation}
G_{1}(x_{j},t;x_{i},0)=\sqrt{\frac{m}{2\pi i\hbar
t}}\exp\left[\frac{im(x_{j}-x_{i})^{2}}{2\hbar t}\right],
\end{equation}

\begin{equation}
G_{2}(x,t+\tau;x_{j},t)=\sqrt{\frac{m}{2\pi
i\hbar\tau}}\exp\left[\frac{im(x-x_{j})^{2}}{2\hbar\tau}\right],
\end{equation}
\begin{equation}
F(x_{j}\pm
d/2)=\frac{1}{\sqrt{\beta\sqrt{\pi}}}\exp\left[-\frac{(x_{j}\pm
d/2)^{2}}{2\beta^{2}}\right],
\end{equation}
and
\begin{equation}
\psi_{0}(x_{i})=\frac{1}{\sqrt{\beta\sqrt{\pi}}}\exp\left(-\frac{x_{i}^{2}}{2\sigma_{0}^{2}}\right).
\end{equation}
The kernels $G_{1}(x_{j},t;x_{i},0)$ and $G_{2}(x,t+\tau;x_{j},t)$
are the free propagators for the particle, the functions $F(x_{j}\pm
d/2)$ describe the double-slit apertures which are taken to be
Gaussian of width $\beta$ separated by a distance $d$; $\sigma_{0}$
is the transverse width of the first slit, where the packet was
prepared, $m$ is the mass of the particle, $t$ ($\tau$) is the time
of flight from the first slit (double-slit) to the double-slit
(screen). We will also consider that the energy associated with the
momentum of the atoms in the $z$-direction is very high, such that
we can consider a classical movement of atoms in this direction,
with the time component given by $z/v_{z}$. This model is presented
in Fig. 1 together with a qualitative illustration of the
interference pattern for three different values of time $t$,
maintaining $\tau$ constant.

\begin{figure}[htp]
\centering
\includegraphics[width=6.0 cm]{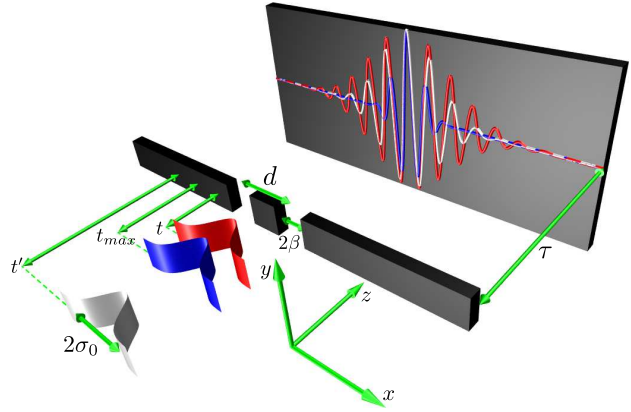}
\caption{Sketch of double-slit experiment. Gaussian wavepacket of
transverse width $\sigma_{0}$ propagates a time $t$ before to attain
the double-slit and a time $\tau$ from the double-slit to the
screen. The slit aperture are taken to be Gaussian of width $\beta$
and separated by a distance $d$. We also show the qualitative
interference pattern considering that the wavepacket propagates the
time $t$ (color red), $t_{max}$ (color blue) or $t^{\prime}$ (color
purple). For $t_{max}$ we have the minimum of interference
fringes.}\label{Figure1}
\end{figure}

After some algebraic manipulations, we obtain for the wave that
passed though the slit $1$ the following result
\begin{equation}
\psi_{1}(x,t,\tau) = \frac{1}{\sqrt{B\sqrt{\pi}}}\exp
\left[-\frac{(x+D/2)^{2}}{2B^{2}}\right]\exp
\left(\frac{imx^2}{2\hbar R} + i\Delta x + i\theta+ i\mu\right),
\end{equation}
where

\begin{equation}
B^{2}(t,\tau)
=\frac{\left(\frac{1}{\beta^{2}}+\frac{1}{b^{2}}\right)^{2}+\frac{m^{2}}{\hbar^{2}}\left(\frac{1}{\tau}+\frac{1}{r}\right)^{2}}
{\left(\frac{m}{\hbar\tau}\right)^{2}\left(\frac{1}{\beta^{2}}+\frac{1}{b^{2}}\right)},
\end{equation}

\begin{equation}
R(t,\tau)=\tau\frac{\left(\frac{1}{\beta^{2}}+\frac{1}{b^{2}}\right)^{2}+\frac{m^{2}}{\hbar^{2}}\left(\frac{1}{\tau}+\frac{1}{r}\right)^{2}}
{\left(\frac{1}{\beta^{2}}+\frac{1}{b^{2}}\right)^{2}+\frac{t}{\sigma_{0}^{2}b^{2}}\left(\frac{1}{\tau}+\frac{1}{r}\right)},
\end{equation}

\begin{equation}
\Delta(t,\tau) = \dfrac{\tau\sigma_{0}^{2}d}{2\tau_0\beta^{2}B^{2}},
\end{equation}

\begin{equation}
D(t,\tau)=\frac{\left(1+\frac{\tau}{r}\right)}{\left(1+\frac{\beta^{2}}{b^{2}}\right)}d,
\end{equation}

\begin{equation}
\theta(t,\tau)=\frac{md^{2}\left(\frac{1}{\tau}+\frac{1}{r}\right)}
{8\hbar
\beta^{4}\left[\left(\frac{1}{\beta^{2}}+\frac{1}{b^{2}}\right)^{2}+\frac{m^{2}}{\hbar^{2}}\left(\frac{1}{\tau}+\frac{1}{r}\right)^{2}\right]},
\end{equation}

\begin{equation}
\mu(t,\tau) =
-\dfrac{1}{2}\arctan\left[\frac{(\frac{t}{\tau_{0}})+\frac{1}{m}(\frac{\hbar
\tau
r}{\tau+r})(\frac{1}{\beta^{2}}+\frac{1}{b^{2}})}{1-\frac{1}{m}(\frac{t}{\tau_{0}})(\frac{\hbar
\tau r}{\tau+r})(\frac{1}{\beta^{2}}+\frac{1}{b^{2}})}\right],
\end{equation}

\begin{equation}
b^2(t) = \sigma_{0}^{2}\left[ 1 + \left(\dfrac{t}{\tau_0}\right)^2
\right],
\end{equation}
and

\begin{equation}
r(t)=t\left[1+\left(\frac{\tau_{0}}{t}\right)^{2}\right].
\end{equation}
In order to obtain the expressions for the wave passing through the
slit $2$, we just have to substitute the parameter $d$ by $-d$ in
the expressions corresponding to the wave passing through the first
slit. Here, the parameter $B(t,\tau)$ is the beam width for the
propagation through one slit, $R(t,\tau)$ is the radius of curvature
of the wavefronts for the propagation through one slit, $b(t)$ is
the beam width for the free propagation and $r(t)$ is the radius of
curvature of the wavefronts for the free propagation. $D(t,\tau)$ is
the separation between the wavepackets produced in the double-slit.
$\Delta(t,\tau)x$ is a phase which varies linearly with the
transverse coordinate. $\theta(t,\tau)$ and $\mu(t,\tau)$ are time
dependent phases and they are relevant only if the slits have
different widths. $\mu(t,\tau)$ is the Gouy phase for the
propagation through one slit. The knowledge of how this phase
depends on time, and particularly on the slit width, can provide us
with some understanding in new designing of double-slit experiment
with matter waves. $\tau_{0}=m\sigma_{0}^{2}/\hbar$ is one intrinsic
time scale which essentially corresponds to the time during which a
distance of the order of the wavepacket extension is traversed with
a speed corresponding to the dispersion in velocity. It is viewed as
a characteristic time for the ``aging" of the initial state
\cite{Carol}.

\section{Phase of the Wavefunction and Position-Momentum Correlations}

In this section we calculate the position-momentum correlations
$\sigma_{xp}$ at the screen and study how they behave as a function of
the propagation times $t$ and $\tau$. We find out that the correlations
present a point of maximum for the propagation time from the source
to the double-slit $t$ whose value depends on $\tau$, the
propagation time from the double-slit to the screen. This point of
maximum express one instability of the phases of the wave function,
which we can associate with incoherence and lack of interference.
Also, we find that the higher the correlations are, the smaller the
region of overlap between the packets sent from each slit will be, i.e., the
maximum of the correlations is associated with a maximum separation
between the two wavepackets when they arrive at the screen.

The normalized wavefunction at the screen is given by
\begin{equation}
\psi(x,t,\tau) = \frac{\psi_{1}(x,t,\tau) +
\psi_{2}(x,t,\tau)}{\sqrt{2+2\exp[-(\frac{D}{2B})^{2}-(\Delta
B)^{2}]}}. \label{psitotal}
\end{equation}
The state \eqref{psitotal} is a superposition of two Gaussians and
therefore presents position-momentum correlations even when $t=0$
\cite{Robinett, Riahi}. For this state we calculate the correlations
and obtain

\begin{eqnarray}
\sigma_{xp}(t,\tau)&=&\frac{1}{2}\langle
\hat{x}\hat{p}+\hat{p}\hat{x}\rangle-\langle \hat{x}\rangle \langle
\hat{p}\rangle\nonumber\\ &=&\frac{mB^{2}}{2R}+\frac{(mD^{2}/R)}{4+4
\exp\left[-\left(\frac{D}{2B}\right)^{2}-\left(\Delta
B\right)^{2}\right]}\nonumber\\
&-&\frac{\hbar\Delta
D}{2}-\frac{(m\Delta^{2}B^{4}/R)}{1+
\exp\left[\left(\frac{D}{2B}\right)^{2}+\left(\Delta
B\right)^{2}\right]}. \label{correlacao}
\end{eqnarray}

We observe that the position-momentum correlations are not dependent
on the terms $\theta$ and $\mu$ and its existence is exclusively due
to the phase dependent of the transverse position $x$. As associated
for the first time by Bohm \cite{Bohm}, the four terms appearing in
the expression for the correlations can be understood as the product
of one ``momentum" by one ``position" for each time $t$ and $\tau$.
For example, the first term is the product of the momentum $(mB/R)$
by the position $B$. The second term is the product of the momentum
$(mD/R)$ by the position $D$. The third term is the product of the
momentum $(\hbar \Delta)$ by the position $D$ and the forth term is
the product of the momentum $(m\Delta^{2}B^{3}/R)$ by the position
$B$. This connection allows us to understand that the higher the
``position" $B$ or $D$ is, the higher the associated ``momentum" and
the contribution to the position-momentum correlations will be. Therefore
this appears as a very simple way to characterize the particle when
it arrives at the screen, allowing us to take a lot of information
about its behavior.

In the following, we plot the curves for the position-momentum
correlations as a function of the times $t$ and $\tau$ for neutrons.
The reason to consider neutrons relies in their experimental
reality, which is most closer to our model for interference with
completely coherent matter waves. We adopt the following parameters:
mass $m=1.67\times10^{-27}\;\mathrm{kg}$, initial width of the
packet $\sigma_{0}=7.8\;\mathrm{\mu m}$ (which corresponds to the
effective width of $2\sqrt{2}\sigma_{0}\approx22\;\mathrm{\mu m}$),
slit width $\beta=7.8\;\mathrm{\mu m}$, separation between the slits
$d=125\;\mathrm{\mu m}$ and de Broglie wavelength
$\lambda=2\;\mathrm{nm}$. These same parameters were used previously
in double-slit experiments with neutrons by A. Zeilinger et al.
\cite{Zeilinger1}. In Fig. 2a, we show the correlations as a
function of $t/\tau_{0}$ for $\tau=18\tau_{0}$, where we observe the
existence of a point of maximum. In Fig. 2b, we show the absolute
value of each term from equation \eqref{correlacao} as a function of
$t/\tau_{0}$ for $\tau=18\tau_{0}$, where we see that the larger
contribution for the position-momentum correlations comes from the
second term, which is directly dependent on the separation
$D(t,\tau)$ between the wavepackets at the screen. Therefore, a
higher separation between the wavepackets at the screen implies
higher position-momentum correlations, i.e., the maximum of the
correlations is associated to a small region of the overlap between
the two packets.

\begin{figure}[htp]
\centering
\includegraphics[width=7.0 cm]{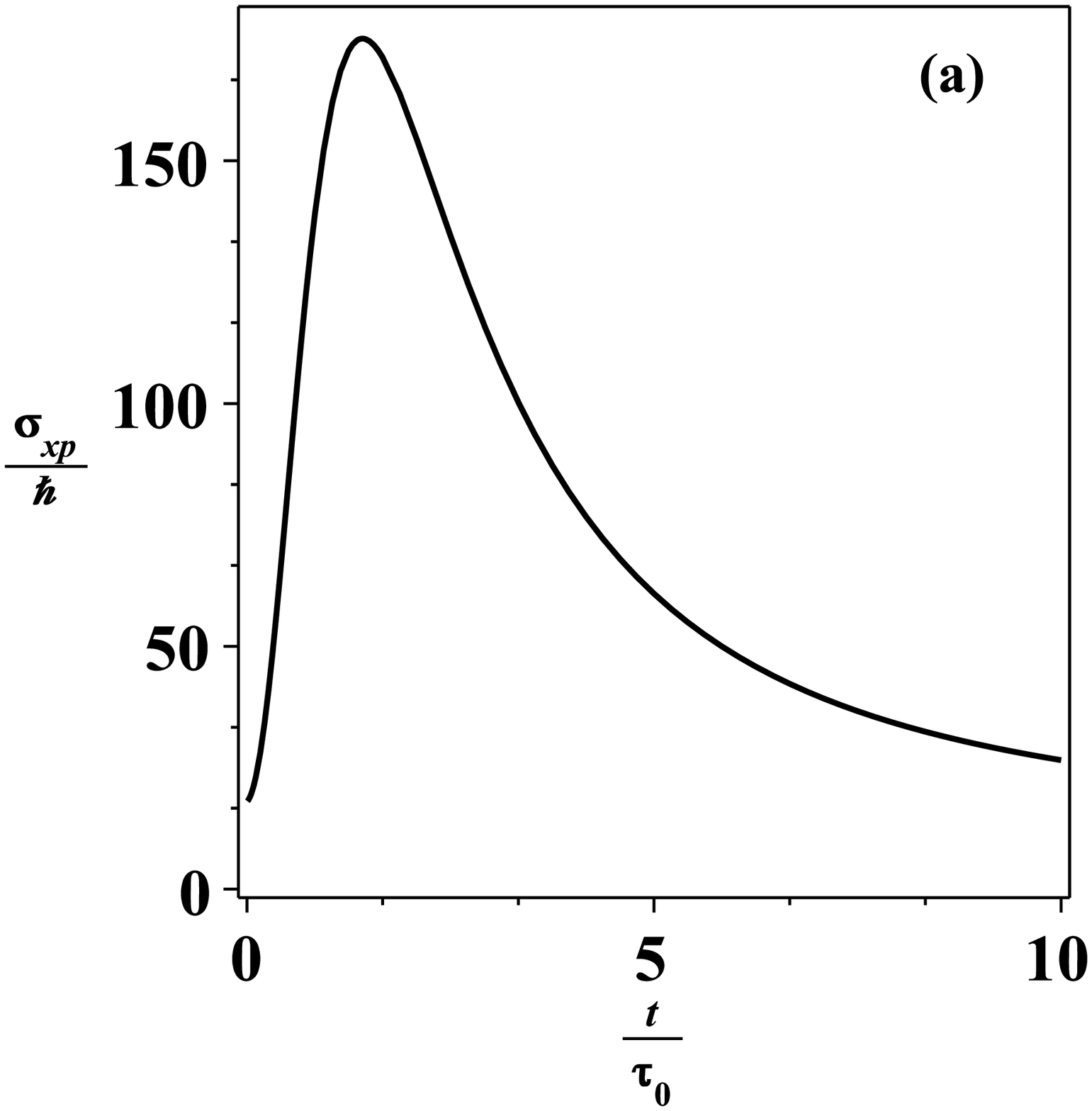}
\includegraphics[width=7.0 cm]{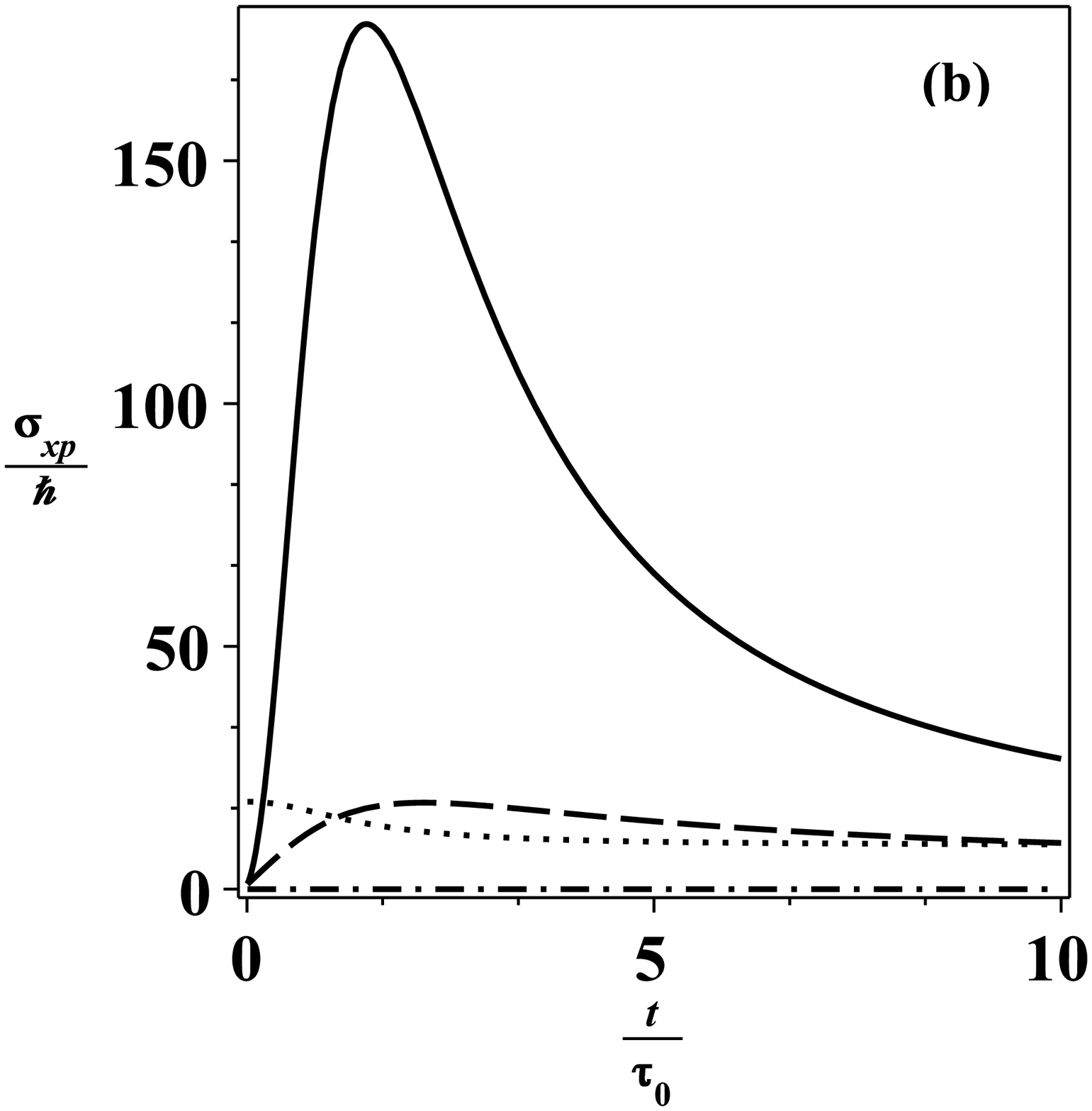}
\caption{(a) Position-momentum correlations as a function of
$t/\tau_{0}$ for $\tau=18\tau_{0}$. (b) Absolute value of the first
(pointed line), second (solid line), third (dashed line) and fourth
(dashed-point line) term of the equation \eqref{correlacao} as a
function of $t/\tau_{0}$ for $\tau=18\tau_{0}$. We observe a point
of maximum and that the larger contribution to the correlations
comes from the second term (solid line) of equation
\eqref{correlacao}.}\label{Figure2}
\end{figure}

In Fig. 3, we show the position-momentum correlations as a function
of $t/\tau_{0}$ and $\tau/\tau_{0}$. We observe that the region
around the point of maximum, or region of phase instability, tends to stay narrower when the propagation time from
the double-slit to the screen $\tau$ increases. We also observe that
the point of maximum is displaced from the left when $\tau$
increases. In the next section we will show a table in which we clearly
see the dependence of the time for the maximum of the correlations
$t_{max}$ with the value of $\tau$, i.e., $t_{max}=t_{max}(\tau)$.
Therefore, the dynamics after the double-slit also influences the
interference pattern and should be taken into account in the
analysis of double-slit experiments. Taking into account only the dynamics before the
double-slit is not sufficient to obtain all the information about the
interference pattern on the screen.

\begin{figure}[htp]
\centering
\includegraphics[width=8.0 cm]{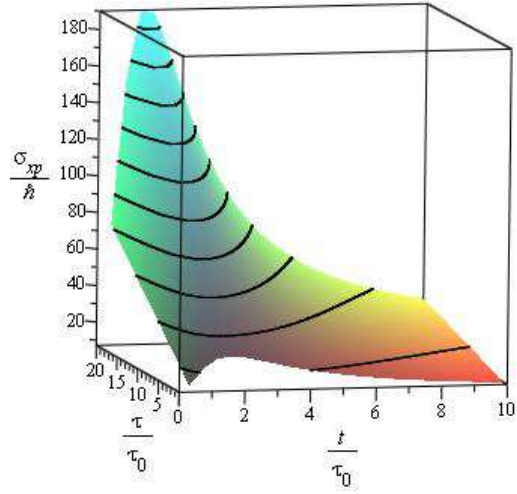}
\caption{Position-momentum correlation as a function of $t/\tau_{0}$
and $\tau/\tau_{0}$. The maximum is displaced to the left and the
region around it tends to stay more narrow when $\tau$ increases.
}\label{Figure3}
\end{figure}

\section{Schrodinger Uncertainty Relation}
It is known that the uncorrelated free particle Gaussian wavepackets
are states of minimum uncertainty both in position and in momentum.
For this case the position-momentum correlations appear only with
the time evolution and are followed by a spreading of the associated
position distribution, while the momentum uncertainty is maintained
constant for all time. For the most general Gaussian wavepacket, in
which the initial position-momentum correlations are present, the
uncertainty in position is minimum at $t=0$ but this is not true for
the uncertainty in momentum \cite{Riahi}. Therefore, the
position-momentum correlations indicate that the uncertainty in one
or in both the quadratures is not a minimum. For the problem treated
here we have a superposition of two Gaussian wavepackets at the
screen, for which the position-momentum correlations are present
indicating that the uncertainty in both the quadratures is not
minimum. To study the behavior of the correlations together with the
behavior of the uncertainties in position and in momentum, we
calculate in this section the determinant of the covariance matrix
defined by

\be\label{Mc} M_C=  \left(\begin{array}{cc}
  \sigma_{xx}^{2} & \sigma_{xp} \\
  \sigma_{xp} & \sigma_{pp}^{2}
\end{array}\right),\ee
where $\sigma_{xx}^{2}=\langle \hat{x}^{2}\rangle-\langle
\hat{x}\rangle^{2}$, $\sigma_{pp}^{2}=\langle
\hat{p}^{2}\rangle-\langle \hat{p}\rangle^{2}$ are the squared
variances in position and momentum, respectively, and $\sigma_{xp}$
is the position-momentum correlations. The expression for
$\sigma_{xp}$ was obtained previously in equation \eqref{correlacao}
and for the other quantities we obtain the following results
\begin{equation}
\sigma^{2}_{xx}(t,\tau)=\frac{B^{2}}{2}+\frac{D^{2}-4\Delta^{2}B^{4}\exp\left[-\left(\frac{D}{2B}\right)^{2}-\left(\Delta
B\right)^{2}\right]}{4+4\exp\left[-\left(\frac{D}{2B}\right)^{2}-\left(\Delta
B\right)^{2}\right]},
\end{equation}
and
\begin{eqnarray}
\frac{\sigma^{2}_{pp}(t,\tau)}{\hbar^{2}}&=&\left(\frac{1}{2B^{2}}+\frac{m^{2}B^{2}}{2\hbar^{2}
R^{2}}\right)+\frac{\left(\frac{mD}{\hbar
R}-2\Delta\right)^{2}}{4+4\exp\left[-\left(\frac{D}{2B}\right)^{2}-\left(\Delta
B\right)^{2}\right]}\nonumber\\
&-&\frac{\left[\frac{D^{2}}{B^{4}}+2\Delta\left(\Delta+\frac{mD}{\hbar
R}\right)\right]}{1+\exp\left[\left(\frac{D}{2B}\right)^{2}+\left(\Delta
B\right)^{2}\right]}.
\end{eqnarray}

The determinant of the covariance matrix, equation (\ref{Mc}), is
the generalized Robertson-Schr\"odinger uncertainty relation and it is
given by
\begin{equation}
D_{C}=\sigma_{xx}^{2}\sigma_{pp}^{2}-\sigma_{xp}^{2}.
\end{equation}
In Fig. 4a we show the curves of the uncertainties $\sigma_{xx}$,
$\sigma_{pp}$ and the correlations $\sigma_{xp}$ normalized to the
same scale as a function of $t/\tau_{0}$ for $\tau=18\tau_{0}$ and
in Fig. 4b we show the determinant $D_{C}/\hbar^{2}$ (solid line) as
a function of $t/\tau_{0}$ for $\tau=18\tau_{0}$, where we compared
it with the value $1/4$ (dashed line). As the position-momentum
correlations mean that both uncertainties are not minima, we see
that this behavior is manifested in the determinant as a fast
increasing in the region around the maximum of the correlations. The
point of maximum is located between the maxima of the uncertainties
in position and in momentum, and the region in which we can consider
the correlations as maximum cover the interval
$0.53\tau_{0}<t<4\tau_{0}$, where $t=0.52\tau_{0}$ is the inflexion
point of the curve of $\sigma_{xp}$ and the other extreme
$t\approx4\tau_{0}$ corresponds to the point for which the
correlations have the same value when $t=0.52\tau_{0}$, i.e.,
$\sigma_{xp}(t=0.52\tau_{0})\approx\sigma_{xp}(t=4\tau_{0})\approx82\hbar$.
On the other hand, the determinant varies slowly in the regions
where the correlations tend to be minima, more specifically the
regions $0<t<0.52\tau_{0}$ and $t>4\tau_{0}$. At the interval
$0<t<0.52\tau_{0}$ the uncertainty in position and in momentum
increases practically by the same rate and at the interval
$t>4\tau_{0}$ the uncertainty in position decreases more slowly than
the uncertainty in momentum. The determinant tends to a constant
value in both intervals, but at the first interval,
$0<t<0.52\tau_{0}$, the curve of correlations has a concavity
upwards in which the value of the determinant tends to the minimum
value $D_{C}\approx16\hbar^{2}$. At the second interval,
$t>4\tau_{0}$, the curve of correlations has a concavity turned down
(tending to a constant function for $t\gg t_{max}$) in which the
determinant tends to the maximum value $D_{C}\approx33\hbar^{2}$.
Then, we observe that $D_{C}>\hbar^{2}/4$ for all time. This
characterizes the non-gaussianity of the state \eqref{psitotal},
since for Gaussian states, initially correlated or not, the
generalized Robertson-Schr\"odinger uncertainty relation is constant
and equal to $\hbar^{2}/4$ for all time. Therefore, for states
obtained from the superposition of two Gaussian states, as the case
treated here, the determinant of the covariance matrix is larger
than $\hbar^{2}/4$ for all time and it is practically constant only
for values of time outside the region around which the correlations
have a point of maximum, showing that the Gaussian features are
strictly altered by the evolution of the position-momentum
correlations. Thus, if we construct one state that has correlations
with a point of minimum, for which the determinant can tend to the
value $\hbar^{2}/4$ at the screen, the number of interference
fringes and its visibility can be increased significantly. It is
possible to do this by considering one double-slit experiment in
which the initial state is the correlated Gaussian state or by
putting a atomic convergent lens next to the double-slit as
similarly has been proposed for light waves \cite{Bartell}.

\begin{figure}[htp]
\centering
\includegraphics[width=7.0 cm]{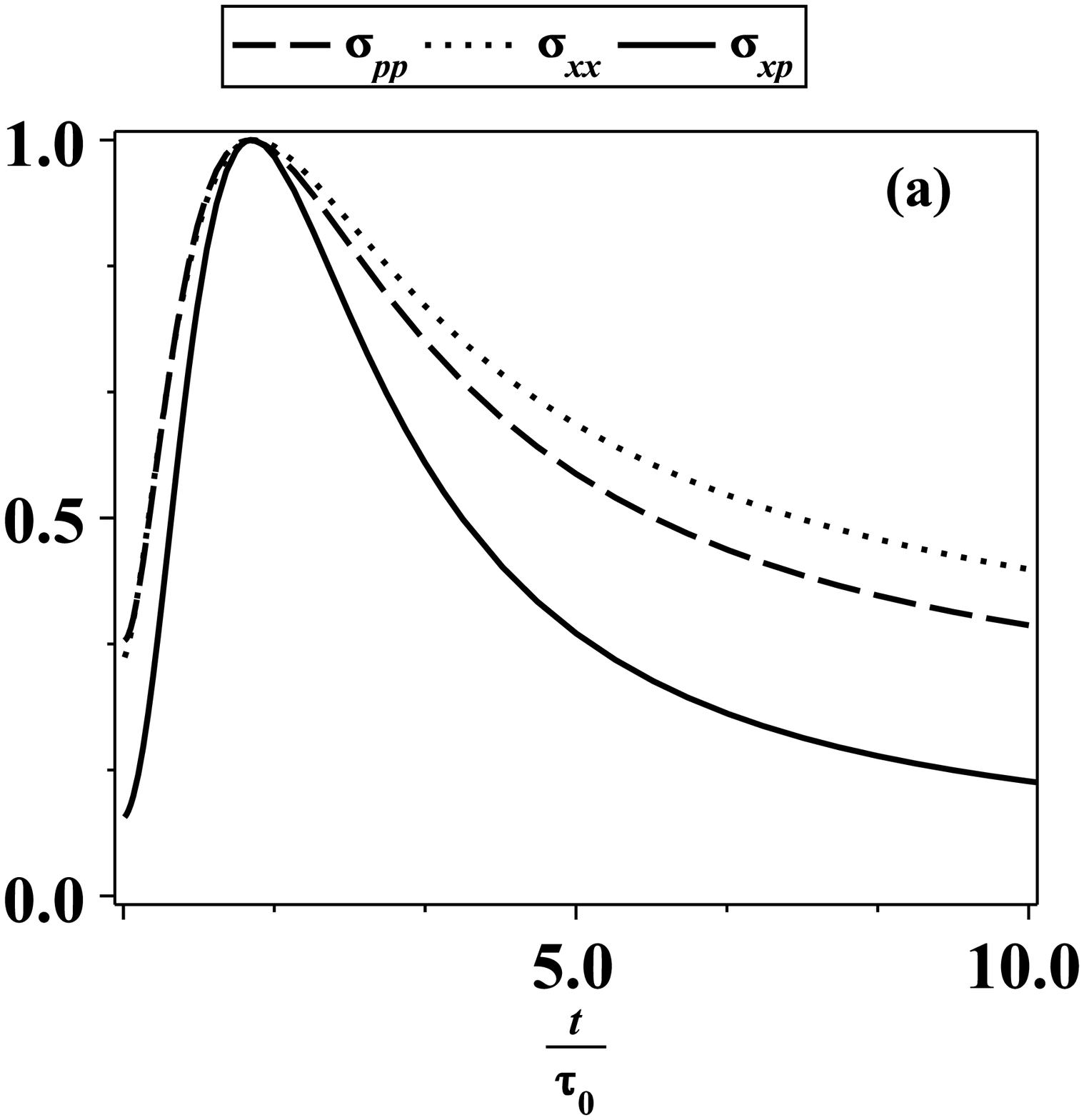}
\includegraphics[width=7.0 cm]{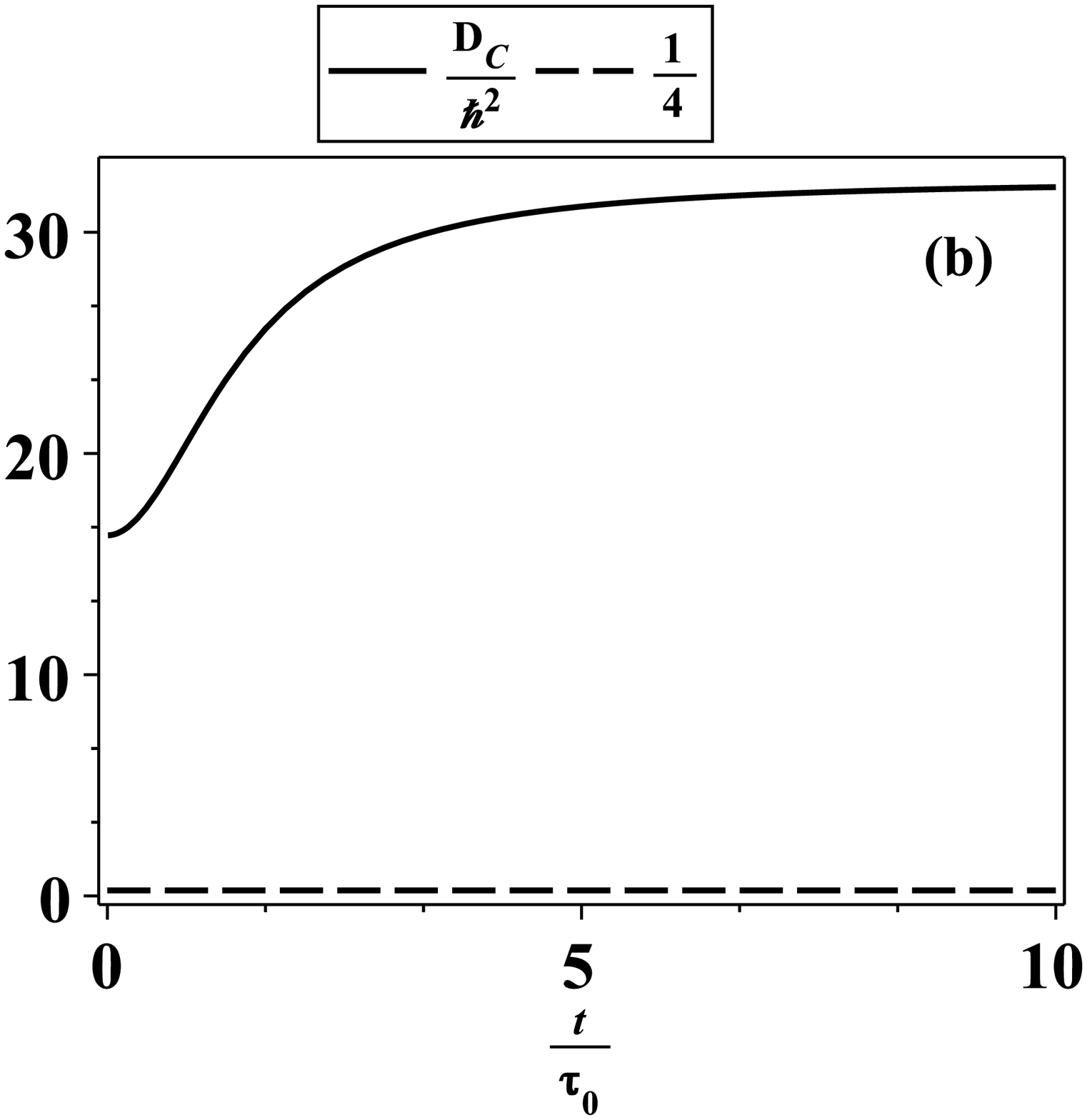}
\caption{(a) Curves of the uncertainties $\sigma_{xx}$,
$\sigma_{pp}$ and the correlations $\sigma_{xp}$ at the same scale
as a function of $t/\tau_{0}$ for $\tau=18\tau_{0}$. (b) Determinant
$D_{C}/\hbar^{2}$ (solid line) as a function of $t/\tau_{0}$ for
$\tau=18\tau_{0}$ compared with the value $1/4$ (dashed line). The
determinant is practically constant at the extremes but different
from the value $\hbar^{2}/4$ and varies rapidly in the region where
the position-momentum correlations have a maximum.}\label{Figure4}
\end{figure}

In table I we show some values of time $t_{max}$ that we calculate
numerically, for which the correlations $\sigma_{xp}$, the
uncertainty in position $\sigma_{xx}$ and the uncertainty in
momentum $\sigma_{pp}$ are maxima and the point of inflexion of the
correlations as a function of time $\tau$. We observe that when
$\tau$ increases, the time $t_{max}$ of the correlations is
dislocated to the left and that this time is always localized
between the times for which the uncertainties in position and in
momentum are maxima. We also observe that the times of maxima tend
to coincide for $\tau>1000\tau_{0}$ and that the time of maximum for
$\sigma_{pp}$ is independent of $\tau$ as a consequence of the free
propagation from the double-slit to the screen.

\begin{table}[ht]
\caption{Times of maxima $t_{max}$ and inflexion $t_{inf}$ as a
function of $\tau$. All terms in units of $\tau_{0}$}
\centering
\begin{tabular}{p{80pt} p{80pt} p{80pt} p{80pt} p{80pt}}
\hline\hline
$\tau$ & $t_{max}$ of $\sigma_{xp}$ & $t_{max}$ of $\sigma_{xx}$ &
$t_{max}$ of $\sigma_{pp}$ & $t_{inf}$ of $\sigma_{xp}$\\ [0.5ex]
\hline
2 & 1.568109061 & 1.984545314 & 1.392356020 &  0.4720349103\\
8 & 1.450312552 & 1.525841616 & 1.392356020 & 0.4990240822\\
18 & 1.419651602 & 1.450522331 & 1.392356020 & 0.5049187153\\
50 & 1.402487095 & 1.413088513 & 1.392356020 & 0.5080737518\\
100  & 1.397465783 & 1.402693625 & 1.392356020 & 0.5089789150\\
1000 & 1.392871030 & 1.393387225 & 1.392356020 & 0.5098004574 \\
[1ex]
\hline
\end{tabular}
\label{table}
\end{table}

\section{Intensity, Visibility and Predictability}
In this section we calculate the relative intensity, visibility and
predictability to analyze the interference pattern, the wave-like
and particle-like behavior from the knowledge of the
position-momentum correlations. Such analysis is very important
because it allows us to choose the set of parameters that provides
the better interference pattern in the double-slit experiment.
The knowledge of the correlations tells us if the particle sent by
the source will behave more as wave-like or particle-like on the
screen. In other words, if the particle is sent by one position for which the
time of flight until the double-slit pertains to the interval around
the maximum of the correlations, it will behave most as a particle for
most values of $x$, excluding only the values near $x=0$.

The intensity on the screen, defined as
$I(x,t,\tau)=|\psi(x,t,\tau)|^{2}$, is given by

\begin{equation}
I(x,t,\tau)=F(x,t,\tau)\left[1+\frac{\cos(2\Delta x)}{\cosh(\frac{D
x}{B^{2}})}\right],
\end{equation}
where

\begin{equation}
F(x,t,\tau)=I_{0}\exp\left[-\frac{x^{2}+(\frac{D}{2})^{2}}{B^{2}}\right]\cosh\left(\frac{D
x}{B^{2}}\right).
\end{equation}

The visibility and predictability are given, respectively, by

\begin{equation}
\mathcal{V}=\frac{I_{max}-I_{min}}{I_{max}+I_{min}}=\frac{1}{\cosh(\frac{Dx}{B^{2}})},
\end{equation}
and
\begin{equation}
\mathcal{P}=\left|\frac{|\psi_{1}|^{2}-|\psi_{2}|^{2}}{|\psi_{1}|^{2}+|\psi_{2}|^{2}}\right|=\left|\tanh\left(\frac{Dx}{B^{2}}\right)\right|.
\end{equation}
The Bohr's complementarity principle established, by the relation of
Greenberger and Yasin for pure quantum mechanical states, that
$\mathcal{P}^{2}+\mathcal{V}^{2}=1$ is satisfied for all values of
$x$ \cite{Greenberger}. The visibility and predictability depend on
the ratio $D/B^{2}$, showing the influence of the parameter $D$ (the
separation between the wavepackets at the time), equivalently the
position-momentum correlations, on the interference pattern.
Therefore, for higher values of $D$ and smaller values of $B$, the
particle-like behavior will be dominant and less visible will be the
interference fringes. As we will see, there is a value of time $t$,
within the interval of maximum correlations, for which the
visibility is minimum and the predictability is maximum. Previously,
the effective number of fringes for light waves in the double-slit
was characterized in Ref. \cite{Bramon} for a given distance (or
time) of propagation from the double-slit to the screen while
neglecting the propagation from the source to the double-slit.
According to \cite{Bramon}, the number of fringes was estimated by a
new index defined by $\nu=0.264/\mathcal{R}$. For the problem
treated here, we have $\mathcal{R}=D/2\Delta B^{2}$, indicating that
the higher the value of $D$ is, the lesser the number of fringes is.

In Fig. 5a, we show the half of the symmetrical plot for the relative
intensity (black line) and in Fig. 5b we show the half of
the symmetrical plot of the visibility (blue line) and predictability
(red line) as a function of $x$ for three different values of $t$,
one of them being the time for which the correlations have a
maximum, with $\tau$ fixed to $\tau=18\tau_{0}$. The corresponding
values of $t$ are, respectively, $t=0.2\tau_{0}$ (solid line),
$t_{max}\approx1.42\tau_{0}$ (dotted line) and $t=18\tau_{0}$
(dashed line). We observe that for $t_{max}\approx1.42\tau_{0}$ the
number of interference fringes is a minimum and the visibility extends
over a small range of the $x$ axis behind the double-slit. In addition, the
predictability dominates extending over a wide range of the $x$ axis. For $t=0.2\tau_{0}$ or $t=18\tau_{0}$ we have a large number of
fringes and the visibility extends over a larger range of the $x$ axis
behind the double-slit. The predictability dominates only in a
range outside the region immediately behind the
double-slit. This shows that a displacement of the source either to the left
or to the right, so that the particles flights a different time from
the times around which the correlations have a maximum $t_{max}$,
most specifically the times in the interval
$0.52\tau_{0}<t<4\tau_{0}$, the number of fringes increases and the
interference pattern presents a better quality. We have to focus on
the region for which the correlations have a maximum and not
specifically at the time of maximum  since although $t_{max}$
really appears as the time for which the number of fringes is a minimum,
the visibility has a minimum in the region of maximum correlations
but it does not coincide with $t_{max}$ being displaced a little from
this point to the right, as we can see in Fig. 6. In fact, for
$t=0.2\tau_{0}$ and $t=18\tau_{0}$ the position-momentum
correlations assume values close to each other, the number of fringes is
nearly the same. However, the visibility is larger for
$t=0.2\tau_{0}$, suggesting that the wave-like
behavior will be most evident when the particle is released closer to the double-slit. Saying in a different way, our ignorance about which slit the particle passed increases when the particle is released closer to the
double-slit. Therefore, although the complementarity relation
$\mathcal{P}^{2}+\mathcal{V}^{2}=1$ is valid for all $x$ independent
of the time (or distance) of propagation, the quantities
$\mathcal{P}(x,t,\tau)$ and $\mathcal{V}(x,t,\tau)$ are
substantially altered at each point $x$ by the propagation times $t$
and $\tau$, as quantitatively shown in Fig. 6.

\begin{figure}[htp]
\centering
\includegraphics[width=7.5 cm]{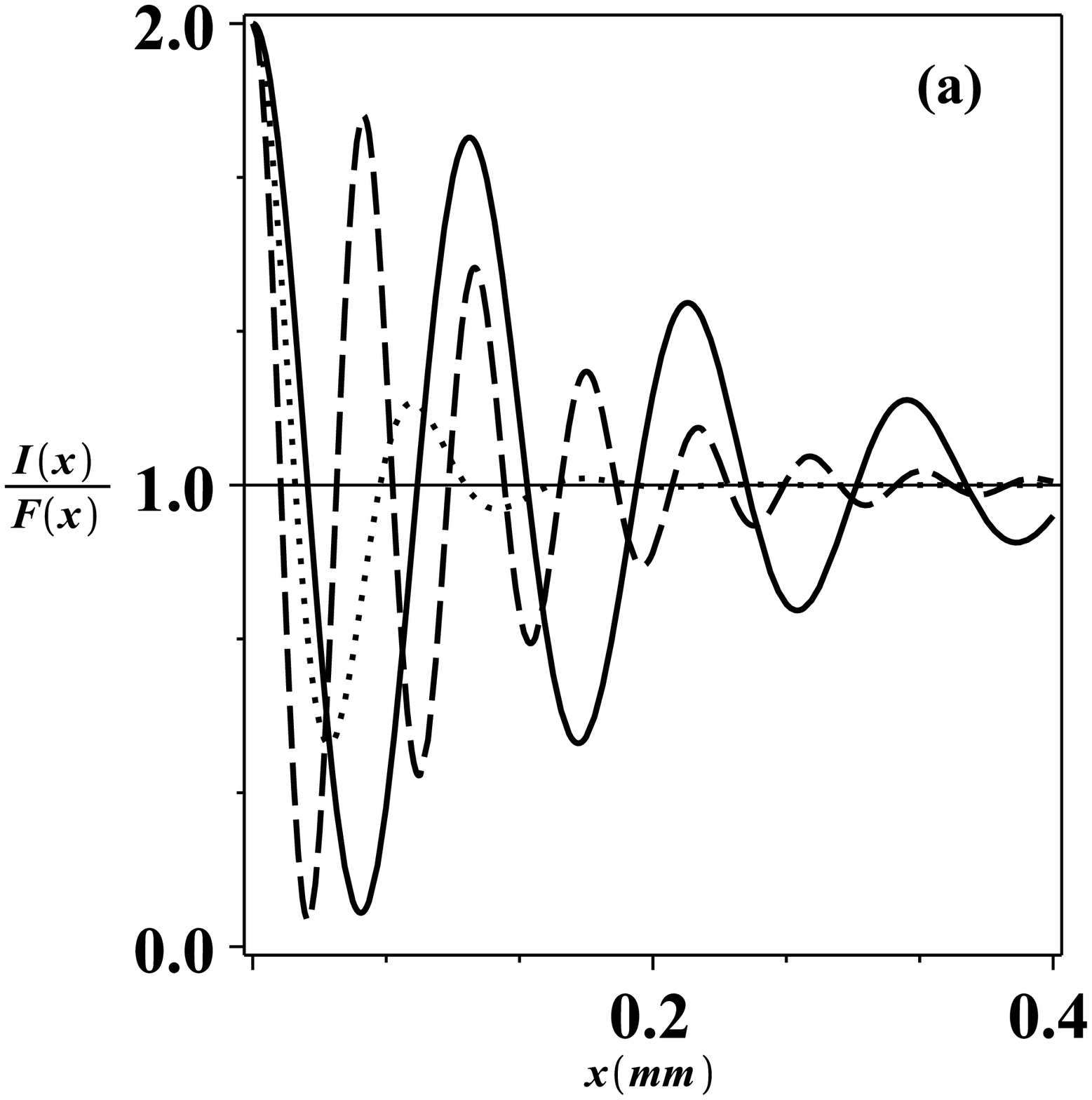}
\includegraphics[width=7.5 cm]{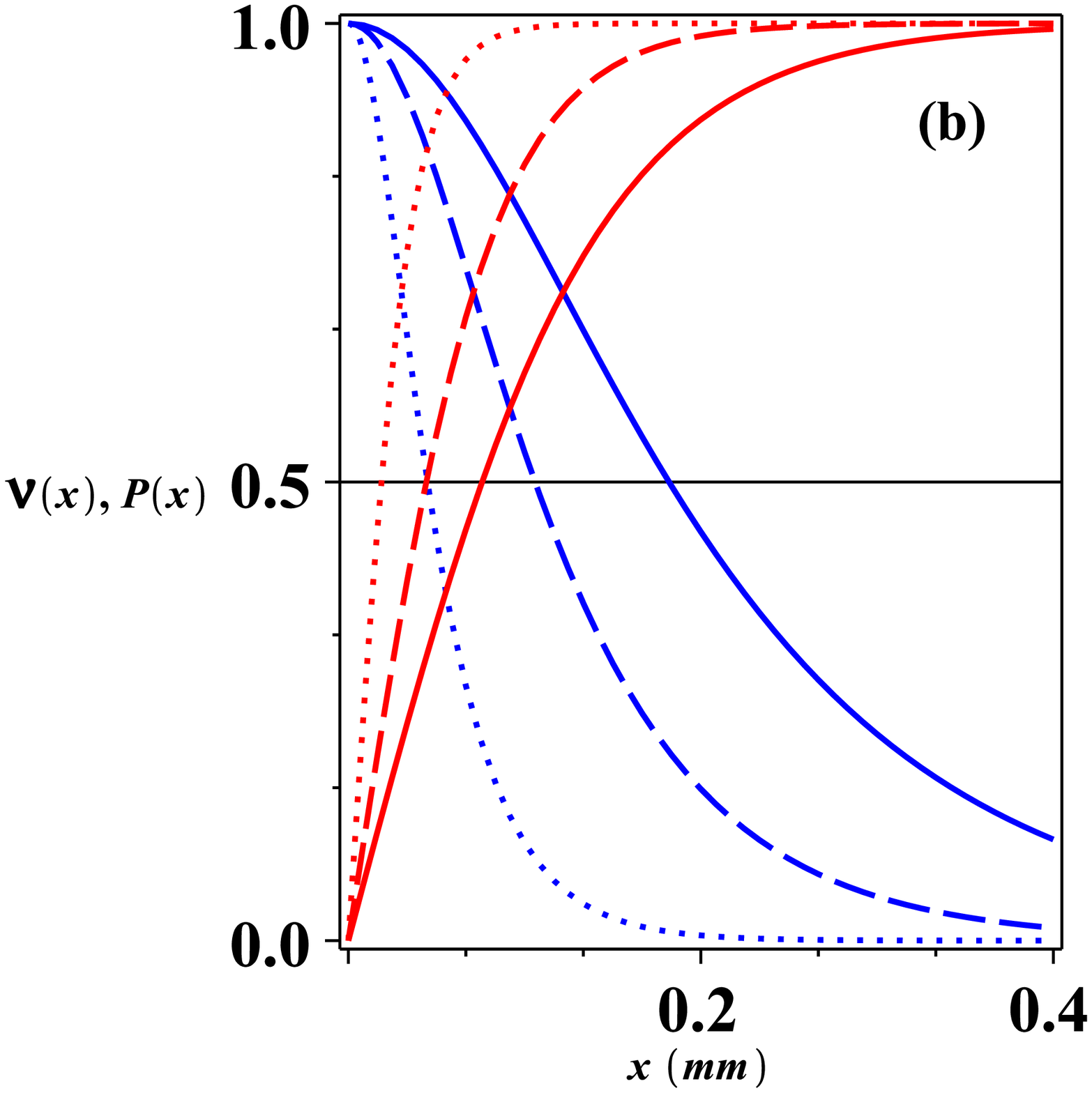}
\caption{(a) Relative intensity (black line) and (b) visibility
(blue line) and predictability (red line) as a function of $x$ for
three different values of $t$ and $\tau=18\tau_{0}$. The
corresponding values of $t$ are, respectively,  $t=0.2\tau_{0}$
(solid line), $t_{max}\approx1.42\tau_{0}$ (dotted line) and
$t=18\tau_{0}$ (dashed line). For these values of time, the time for
which the correlations have a maximum $t_{max}\approx1.42\tau_{0}$
presents the least number of fringes and visibility. Moving the
source of particles to the left or to the right from the region
around the maximum of the correlations, the number of fringes and
visibility increase.}\label{Figure5}
\end{figure}

\begin{figure}[htp]
\centering
\includegraphics[width=5 cm]{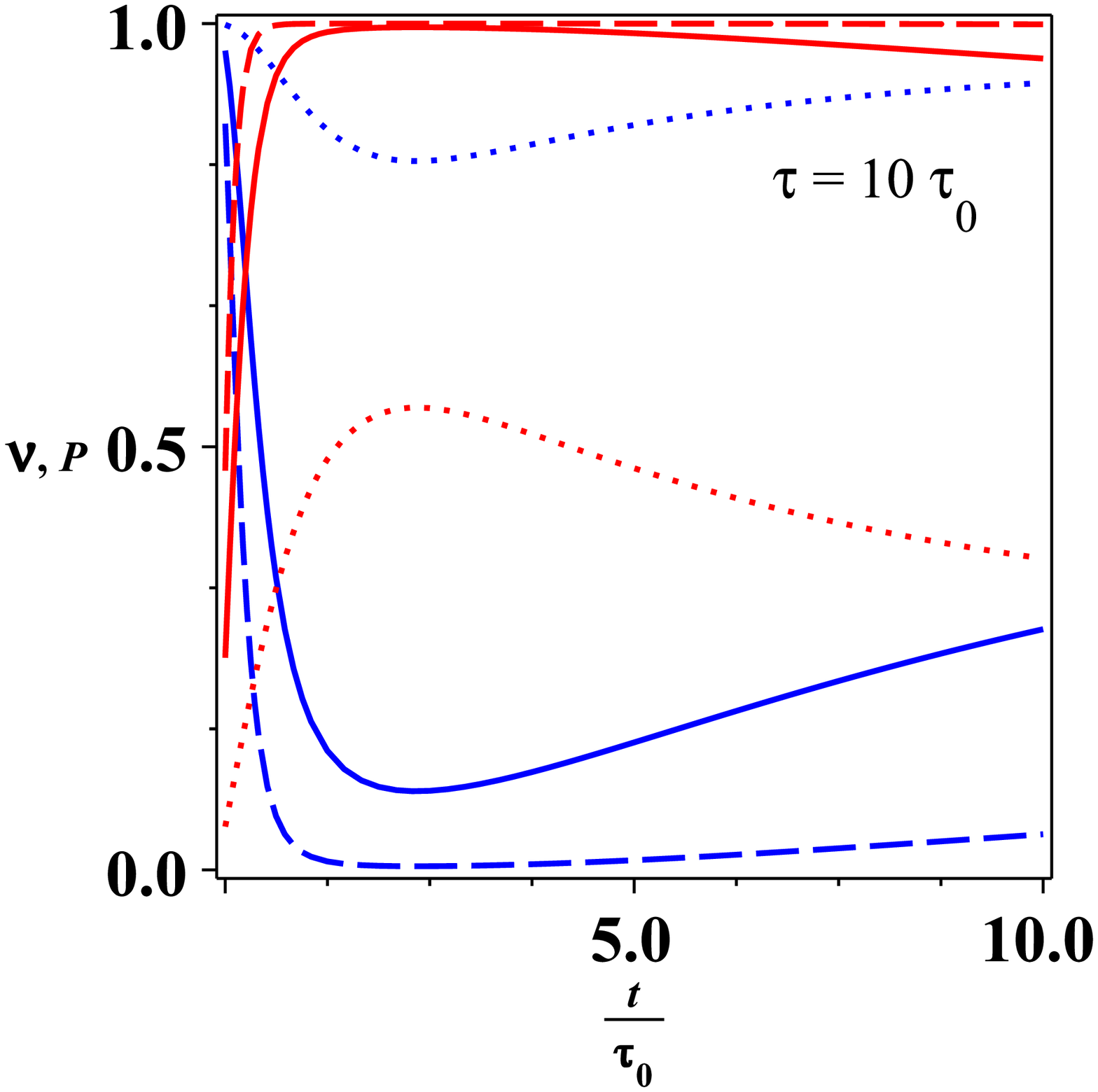}
\includegraphics[width=5 cm]{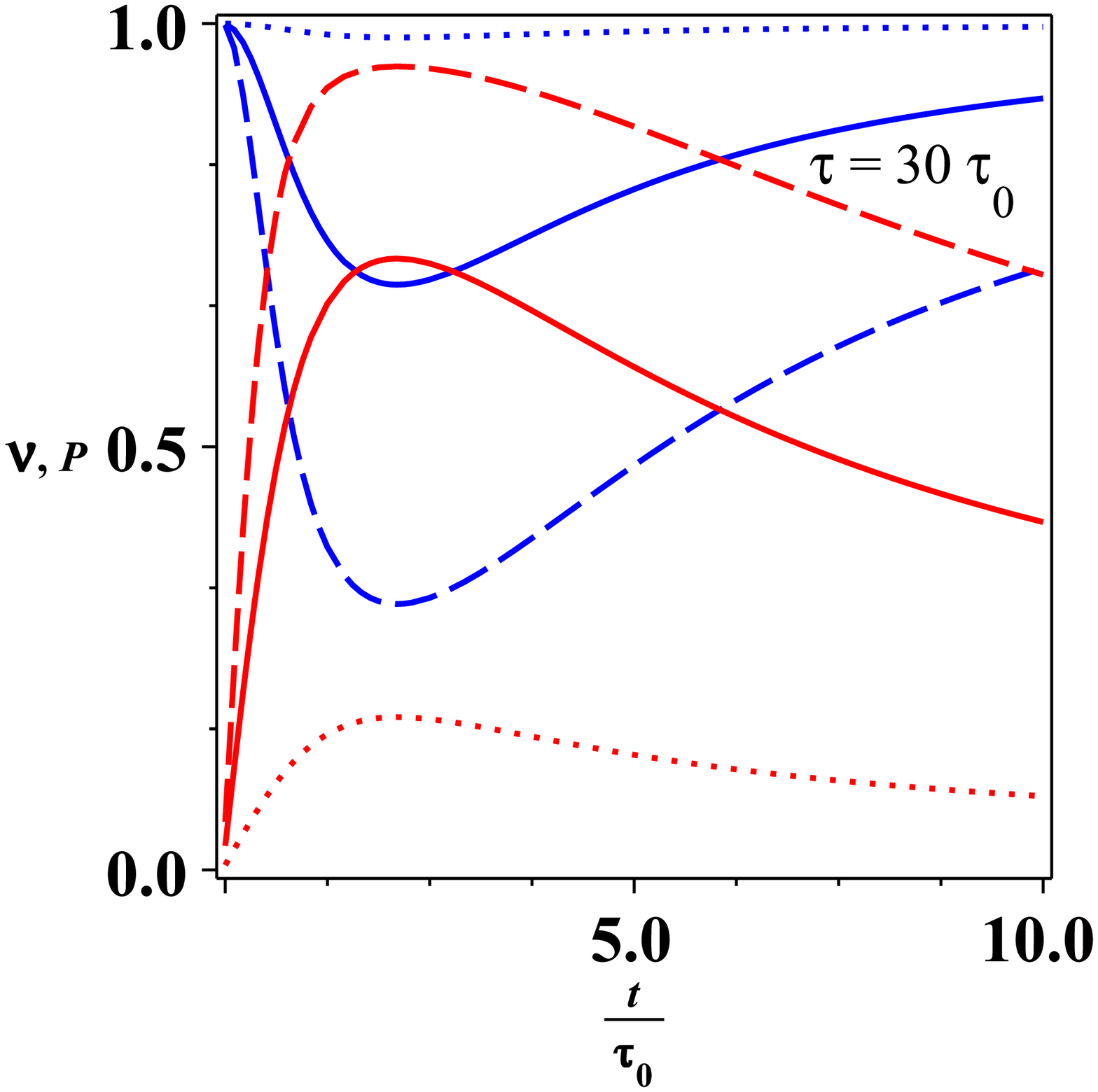}
\includegraphics[width=5 cm]{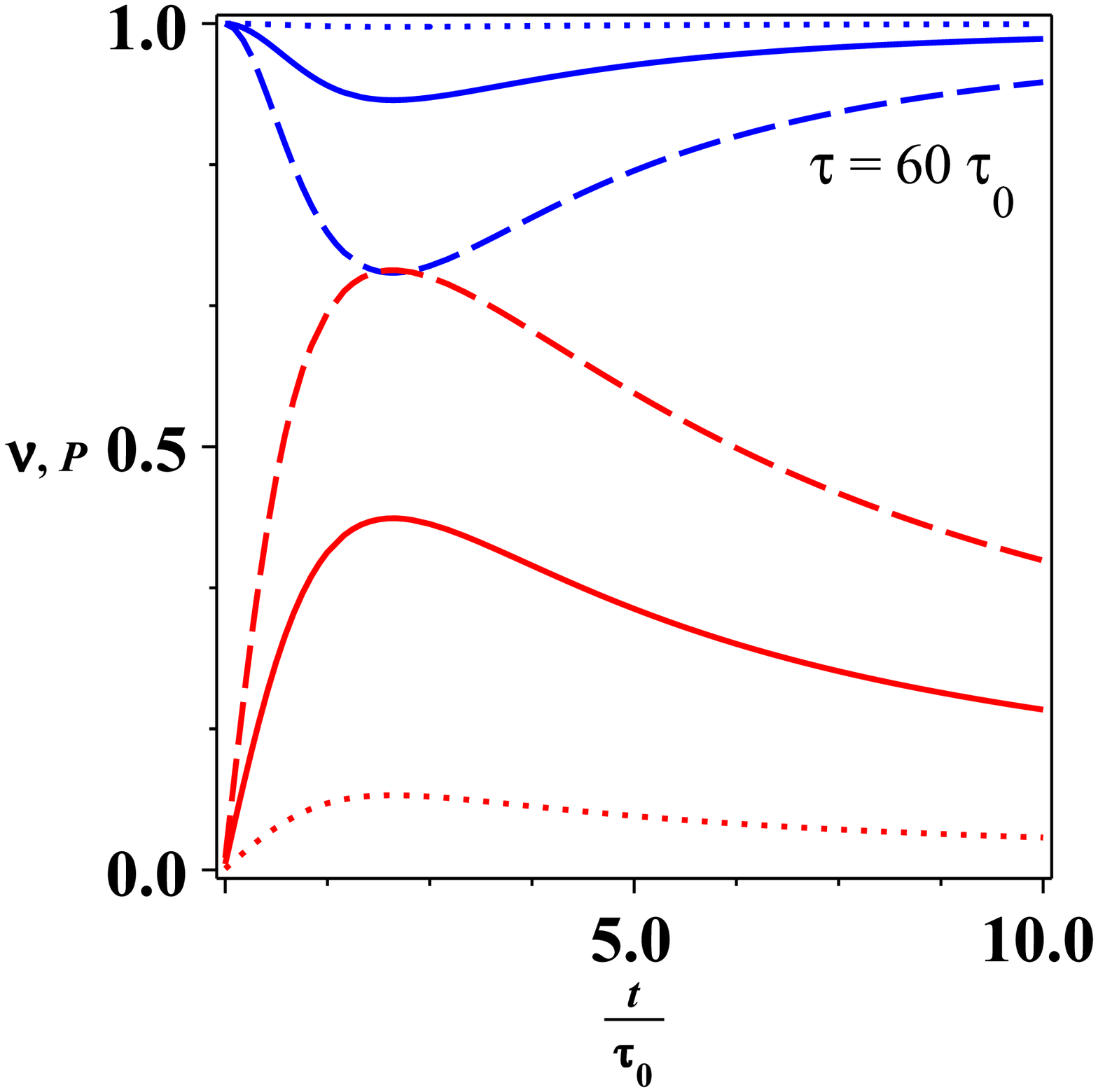}
\caption{Visibility (blue line) and  predictability (red line) as a
function of $t/\tau_{0}$ for three different values of $x$. The
corresponding values of $x$ are $x=0.01\;\mathrm{mm}$ (dotted line),
$x=0.05\;\mathrm{mm}$ (solid line) and $x=0.1\;\mathrm{mm}$ (dashed
line). We present figures for $\tau=10\tau_{0}$, $\tau=30\tau_{0}$
and $\tau=60\tau_{0}$. The values of $\mathcal{V}$ and $\mathcal{P}$
for each value of $x$ are strongly altered by the values of $t$ and
$\tau$. For example, exist a value of time $t$ for which the
visibility is minimum and the predictability is maximum and for
$\tau>60\tau_{0}$ the values of $\mathcal{V}$ are higher than the
values of $\mathcal{P}$.} \label{Figure6}
\end{figure}

In Fig. 7a, we show the half of the symmetrical plot for relative
intensity (black line) and in Fig. 7b we show the half of the
symmetrical plot of the visibility (blue line) and predictability
(red line) as a function of $x$ for two different values of $\tau$,
fixing $t$ at $t=8\tau_{0}$. The corresponding values of $\tau$ are,
respectively, $\tau=10\tau_{0}$ (dashed line) and $\tau=30\tau_{0}$
(solid line). For $\tau=30\tau_{0}$, we have a larger number of
fringes with a better visibility because the region of maximum
correlations will be further for $t=8\tau_{0}$ with
$\tau=30\tau_{0}$ than the $\tau=10\tau_{0}$ case, according to table I.
In this case we observe that the displacement of the maximum of the
correlations implies an increasing in the spatial transverse
coherence with time. In fact, the number of interference fringes is
nearly the same for both values of $\tau$, but the visibility is
larger for $\tau=30\tau_{0}$ in comparison with $\tau=10\tau_{0}$. This
shows that the wave-like behavior becomes more evident,
comparatively, when the particle is launched from a position such
that the flight time until the double-slit is most distant
from the time for which the correlations have a maximum. On the
other hand, we can say that our ignorance about which slit the
particle passed, when it is launched from the position
$z=v_{z}(t=8\tau_{0})$, is smaller when the screen is positioned at
$z_{1}=v_{z}(\tau_{1}=10\tau_{0})$ than the situation where the screen
is positioned at $z_{2}=v_{z}(\tau_{2}=30\tau_{0})$. Again, we see the
influence of the times $t$ and $\tau$ over the quantities
$\mathcal{P}(x,t,\tau)$ and $\mathcal{V}(x,t,\tau)$, although the
result $\mathcal{P}^{2}+\mathcal{V}^{2}=1$ is maintained for all $x$ values
independent of the time.

\begin{figure}[htp]
\centering
\includegraphics[width=7.5 cm]{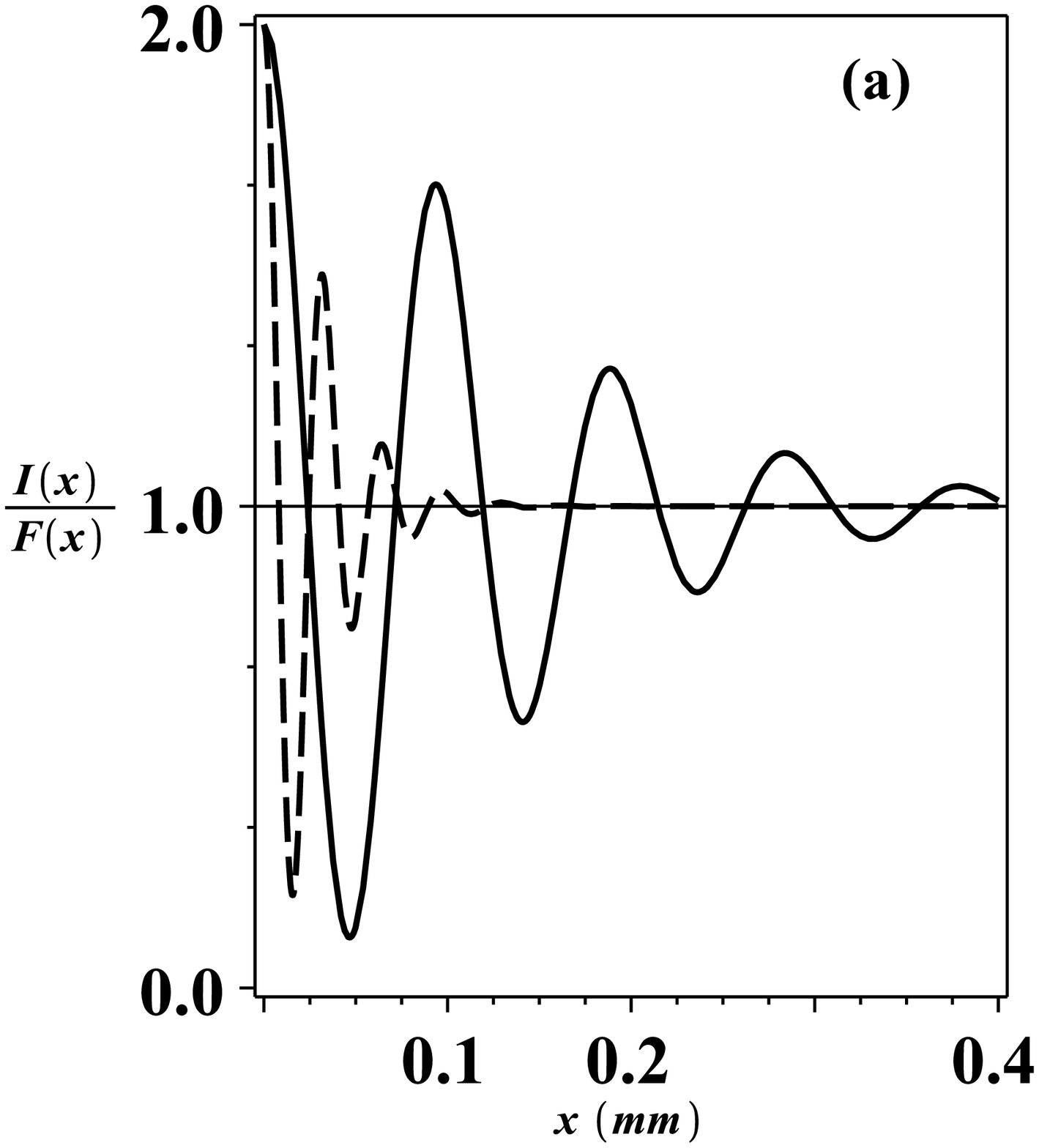}
\includegraphics[width=7.5 cm]{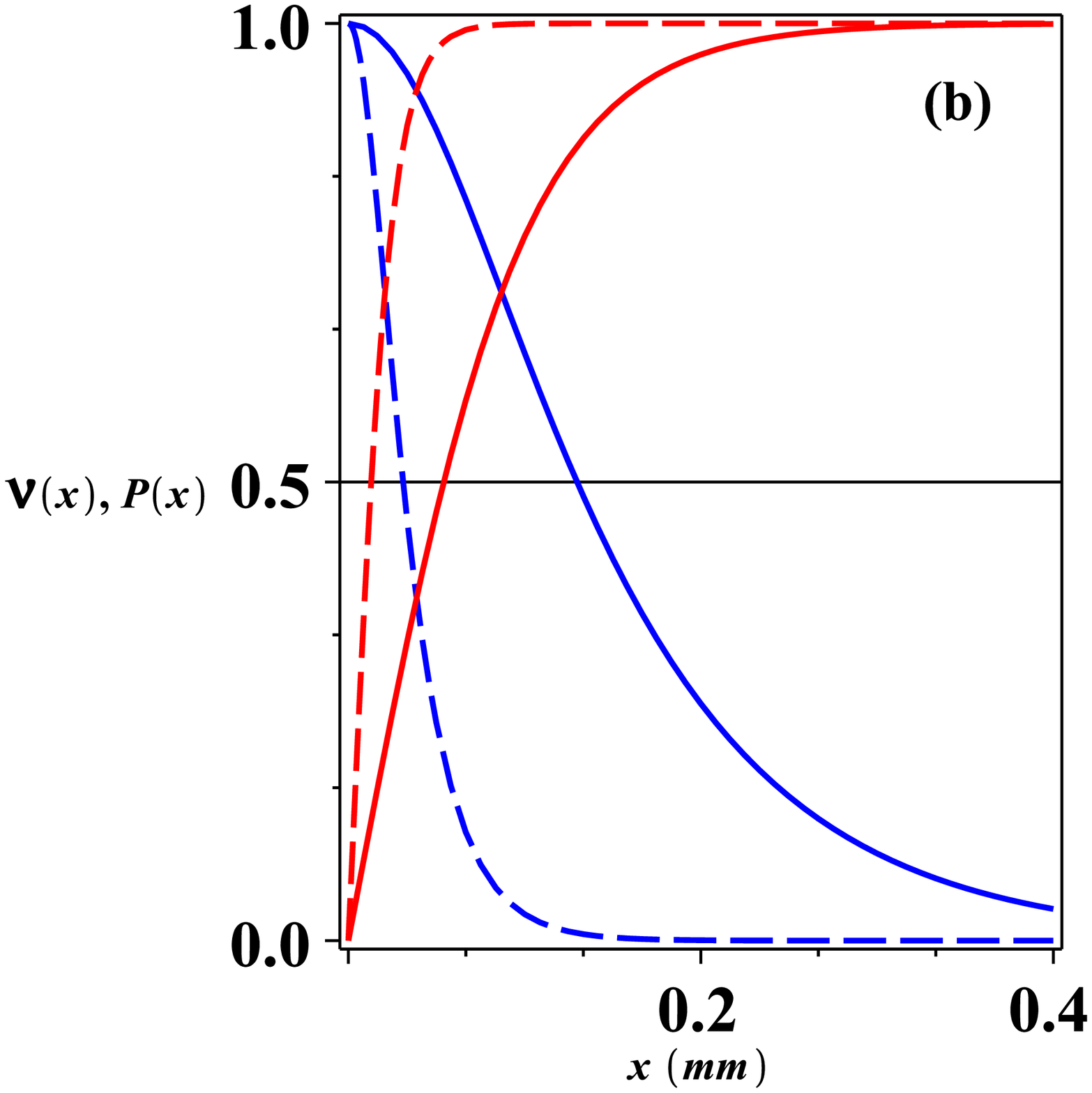}
\caption{(a) Relative intensity (black line) and (b) visibility
(blue line) and predictability (red line) as a function of $x$ for
two different values of $\tau$ and $t$ fixed in $t=8\tau_{0}$. The
corresponding values of $\tau$ are, respectively, $\tau=10\tau_{0}$
(dashed line) and $\tau=30\tau_{0}$ (solid line). For
$\tau=30\tau_{0}$, we have the most number of interference fringes
with a better visibility because the point for which the
correlations have a maximum is more distant of $t=8\tau_{0}$ for
$\tau=30\tau_{0}$, according to table I. In fact, the number of
fringes is practically the same but the visibility is considerably
larger for $\tau=30\tau_{0}$.}\label{Figure7}
\end{figure}

The results above were obtained for neutrons treated as wavepackets of initial transverse width
$\sigma_{0}=7.8\;\mathrm{\mu m}$. For these parameters, the time scale is given by $\tau_{0}=m\sigma_{0}^{2}/\hbar=1.02\;\mathrm{ms}$.
We can note a good quality in the interference pattern for
$t=18\tau_{0}=18.02\;\mathrm{ms}$ and $\tau=18\tau_{0}=18.02\;\mathrm{ms}$, whose velocity around
$v=200\;\mathrm{m/s}$, corresponds to distances $z_{t}=3.6\;\mathrm{m}$ and
$z_{\tau}=3.6\;\mathrm{m}$. These parameters were used by A.
Zeilinger et al. and they correspond to distances within the experimental viability \cite{Zeilinger1}. Now, if we take, for instance, the mass of the order of
$m=1.2\times10^{-24}\;\mathrm{kg}$, which is next to the mass of the fullerene molecules,
and build a package of the same width of the neutrons, we will have $\tau_{0}=0.73\;\mathrm{s}$.
In this case, $t=18\tau_{0}=13.14\;\mathrm{s}$ and $\tau=18\tau_{0}=13.14\;\mathrm{s}$.
Considering one velocity of $200\;\mathrm{m/s}$, we will have
$z_{t}=2.63\times10^{3}\;\mathrm{m}$ and $z_{\tau}=2.63\times10^{3}\;\mathrm{m}$, which are distances outside the experimental reality.
Therefore, by analyzing the behavior of the correlations, we can also capture
information about the difficulty in observing interference with
macroscopic objects.

In Ref. \cite{Carol} the authors explore the effect of the position-momentum correlations on the interference pattern
but they do not take into account the influence of the propagation time from the double-slit to the screen. They also do not discuss the behavior of the correlations as a function of the propagation time from the source to the double-slit
(or equivalently, the behavior of the correlations as a function of the parameter $\sigma_{0}$). We observe that for
the parameters used in this reference, the correlations are maxima for $0.013\;\mathrm{\mu m}\leq\sigma_{0}\leq0.02\;\mathrm{\mu m}$
and minima for $\sigma_{0}>1.0\;\mathrm{\mu m}$, which justify the poor interference pattern for $0.013\;\mathrm{\mu m}\leq\sigma_{0}\leq0.02\;\mathrm{\mu m}$
and a rich interference pattern for $\sigma_{0}=6.0\;\mathrm{\mu m}$.

\section{Conclusions}
In this contribution, we studied the double-slit experiment as an attempt to find parameters that
produce the maximum number of interference fringes and with the highest possible quality on the screen.
Our results show that we can take information about the
interference pattern by looking at the behavior of the position-momentum correlations,
that are installed with the quantum dynamics. We observe that both the dynamics before and after the double-slit are important for the existence and quality of the interference fringes on the screen. Especially we observe that there is a value
of propagation time from the source to the double-slit for which the correlations have a point of maximum,
so that particles released by a source at the region around this point produce interference fringes on the screen with the worst
quality. The wave-like and particle-like behavior expressed by the complementary relation of Greenberger and Yasin
$\mathcal{P}^{2}+\mathcal{V}^{2}=1$ is also strongly influenced at each point $x$ by the times $t$ and $\tau$, i.e., depending where the particle came from
and where the screen was positioned, it will behave most as a wave or most as a particle at the screen.
The knowledge of the point of maximum of the position-momentum correlations can also help us to choose the best
parameters which allow us to observe interference effects with macromolecules, such as fullerenes.
From the determinant of the covariance matrix it was possible to observe how the Gaussian properties of the state produced on the screen by
the superposition of two Gaussian are altered when the uncertainties in position and in momentum and the position-momentum
correlations vary with the times $t$ and $\tau$.

\begin{acknowledgments}
\section*{Acknowledgments}
We would like to thank the Professor E. C. Girão by careful reading
of the manuscript. I. G. da Paz and L. A.Cabral acknowledge useful
discussions with M.C. Nemes. J. S. M. Neto thanks the CAPES by
financial support under grant number 210010114016P3. I. G. da Paz
thanks support from the program PROPESQ (UFPI/PI) under grant number
PROPESQ 23111.011083/2012-27.
\end{acknowledgments}


\end{document}